\documentclass[printer]{aa}
\usepackage{graphicx}
\usepackage{natbib}

\newcommand{\xiv}{\mbox{\boldmath$\xi$}}
\def\lsim{\mathrel{\rlap{\lower4pt\hbox{\hskip1pt$\sim$}}
    \raise1pt\hbox{$<$}}}                
\def\gsim{\mathrel{\rlap{\lower4pt\hbox{$\sim$}}
    \raise1pt\hbox{$>$}}}                
\begin{document}
\authorrunning{Reisenegger}
\titlerunning{Magnetic equilibria in stars}
   \title{Stable magnetic equilibria and their evolution in the\\
    upper main sequence, white dwarfs, and neutron stars}


   \author{Andreas Reisenegger\inst{1,2}          }

   \offprints{A. Reisenegger}

   \institute{Departamento de Astronom{\'\i}a y Astrof{\'\i}sica,
             Pontificia Universidad Cat\'olica de Chile, Casilla 306, Santiago 22,
             Chile\thanks{Permanent address}\\
              \email{areisene@astro.puc.cl}\\
         \and
             Max-Planck-Institut f\"ur Astrophysik, Karl-Schwarzschild-Str. 1,
             85741 Garching bei M\"unchen, Germany
             }

   \date{Received ; accepted }

  \abstract
   {Long-lived, large-scale magnetic field configurations exist in upper main
   sequence, white dwarf, and neutron stars. Externally, these fields
   have a strong dipolar component,
   while their internal structure and evolution are uncertain
   but highly relevant to several problems in
   stellar and high-energy astrophysics.}
   {We discuss the main properties expected for the
   stable magnetic configurations in these stars from physical arguments and the ways these
   properties may determine the modes of decay of these configurations.}
   {We explain and emphasize the likely importance of the non-barotropic, stable stratification
   of matter
   in all these stars (due to entropy gradients in main-sequence envelopes and white dwarfs,
   due to composition gradients in neutron stars). We first illustrate it in
   a toy model involving a single, azimuthal magnetic flux tube.
   We then discuss the effect of stable stratification or its
   absence on more general configurations, such as axisymmetric equilibria involving poloidal
   and toroidal field components. We argue that the main mode of
   decay for these configurations are processes that lift the
   constraints set by stable stratification, such as heat
   diffusion in main-sequence envelopes and white dwarfs, and beta
   decays or particle diffusion in neutron stars. We estimate the
   time scales for these processes, as well as their interplay with
   the cooling processes in the case of neutron stars.}
   {Stable magneto-hydrostatic equilibria appear to exist in stars whenever the
   matter in their interior is stably stratified (not barotropic).
   These equilibria are not force-free and not required to satisfy the
   Grad-Shafranov equation, but they do
   involve both toroidal and poloidal field components. In main sequence stars with radiative
   envelopes and in white dwarfs, heat diffusion is not fast enough to make
   these equilibria evolve over the stellar lifetime. In neutron stars, a strong enough field
   might decay by overcoming the compositional stratification through beta decays
   (at the highest field strengths) or through ambipolar diffusion (for somewhat weaker fields).
   These processes convert magnetic
   energy to thermal energy, and they occur at significant rates only once the latter is
   less than the former; therefore, they substantially delay the cooling of the neutron
   star, while slowly decreasing its magnetic energy.}
   {}


   \maketitle
%

\section{Introduction}
\label{sec:intro}

   Upper main sequence stars, white dwarfs, and neutron stars appear
   to have long-lived magnetic fields. These fields are organized on
   large scales, in the sense that
   the dipole field components (and perhaps some other, low-order
   multipoles; e.~g. \citealt{Bagnulo99,Bagnulo00}) are not much weaker than
   the rms surface field, unlike the highly chaotic field of the Sun.

   The highest detected (surface dipole) magnetic field strengths are
   $B_\mathrm{max}\sim 10^3~\mathrm{G}$ in O
   stars (radius $R\sim 10~R_{\sun}$; \citealt{Donati02,Donati06,Petit08}),
   $B_\mathrm{max}\sim 3\times 10^4~\mathrm{G}$
   in chemicallly peculiar A and B stars (Ap/Bp stars, $R\sim 3~R_{\sun}$;
   \citealt{Mathys97,Bagnulo99}), $B_\mathrm{max}\sim 10^9~\mathrm{G}$ in white dwarfs
   ($R\sim 10^4\mathrm{km}$; \citealt{Schmidt03}), and
   $B_\mathrm{max}\sim 10^{15}~\mathrm{G}$ in ``magnetars'', a subclass of
   strongly magnetized neutron stars ($R\sim 10~\mathrm{km}$;
   \citealt{Kouveliotou98,Woods99}), in all cases yielding very
   similar total magnetic fluxes, $\Phi_\mathrm{max}=\pi R^2 B_\mathrm{max}\sim
   10^{27.5}~\mathrm{G~cm^2}$. This coincidence has often been interpreted as an
   argument for flux freezing during stellar evolution
   \citep{Ruderman72,R01b,R03,Ferrario05a,Ferrario05b,Ferrario06},
   although its feasibility has been called into question
   \citep{TD93,Spruit08}. Of course, a large fraction
   of the original magnetic flux might be ejected with the stellar
   envelope. On the other hand, substantial field amplification through
   differential rotation, convection, and various instabilities could plausibly occur
   in proto-neutron stars if they are born rapidly rotating \citep{TD93,Spruit02,Spruit08}.

   Connected to the similar fluxes is that
   these stars also have similar ratios of fluid pressure ($P\sim
   GM^2/R^4$, where $G$ is the gravitational constant and
   $M$ is the mass of the star) to magnetic pressure ($B^2/8\pi$),
   \begin{equation}\label{beta}
   \beta={8\pi P\over B^2}\sim{8\pi^3GM^2\over\Phi^2}\sim 3\times
   10^6\left(M\over M_{\sun}\right)^2\left(\Phi\over\Phi_\mathrm{max}\right)^{-2},
   \end{equation}
   a large number, even for the most highly magnetized objects, implying that the
   magnetic field causes only very minor perturbations to their hydrostatic equilibrium
   structure.

   Another similarity among these stars is that much or all of their
   structure is stably stratified, i.~e. stable to convection.
   The radiative envelopes of upper main sequence stars, as well
   as the whole interior of white dwarfs, are stabilized by the
   radially increasing specific entropy $s$, while in neutron stars
   the same effect is caused by a radially varying mix of different
   particle species \citep{Pethick92,RG92,R01a}, which in the outer
   core reduces to a radially varying proton and electron
   fraction, $Y\equiv n_p/(n_n+n_p)=n_e/(n_n+n_p)$, where $n_i$
   stands for the number densities of neutrons ($i=n$), protons ($i=p$) and
   electrons ($i=e$).

   The structure of the magnetic field inside these stars is not known, although
   it is highly relevant to their evolution:
   \begin{itemize}
   \item[1)] It affects the radial transport of angular momentum and
   chemical elements (e.~g. \citealt{Heger05});
   \item[2)] it is plausibly the dominant source of energy for both
   the outbursts and the persistent emission of soft gamma repeaters
   (SGRs) and anomalous X-ray pulsars (AXPs), for this reason collectively
   called ``magnetars'' \citep{TD93,TD96};
   \item[3)] it is likely to play an important role in the frequency
   spectrum of quasi-periodic oscillations observed after SGR
   flares \citep{Levin07};
   \item[4)] it leads to slight deformations of neutron stars that could give
   rise to precession
   of pulsars \citep{Wasserman03} and to the emission of gravitational waves
   \citep{Cutler02}.
   \end{itemize}

   Magnetohydrodynamic (MHD) simulations of stably stratified stars with random initial
   magnetic field configurations have shown them to
   evolve on Alfv\'en-like time scales into
   long-lived structures whose further evolution and decay appears to be controlled by
   dissipative processes (in the simulations, Ohmic diffusion;
   \citealt{BS04,BS06,BN06}). Often, but not always \citep{Braithwaite08}, these
   are roughly axisymmetric combinations of linked poloidal and
   toroidal components, whose external appearance is essentially dipolar.
   It appears plausible that these configurations approximate the true magnetic
   field structures in upper main sequence stars, white dwarfs, and neutron stars.

   In \S~2, we present arguments to the effect that the stable
   stratification of the stellar matter should be an essential ingredient to
   these equilibria. This means that, contrary to assumptions in the recent literature
   (e.~g. \citealt{PerezAzorin06,Broderick08,Mastrano08}),
   these are definitely \emph{not} force-free fields. In fact, it
   is shown in Appendix A that there are no true force-free equilibria
   in stars, while those proposed in the literature actually have
   singular magnetic forces on the stellar surface.
   Moreover, the fluid \emph{cannot} be treated as barotropic, therefore the field
   components are \emph{not} required to satisfy the Grad-Shafranov equation
   \citep{Mestel56}, contrary to the popular belief
   \citep{Tomimura05,Yoshida06,Haskell08,Akgun08,Kiuchi08}.
   In fact, the range of available
   equilibria becomes much wider in a stably stratified, non-barotropic
   fluid. The constraints imposed by the stability of these equilibria are far from
   obvious, but we argue that there are probably no equilibria in barotropic stars,
   while it is likely that there are equilibria with linked toroidal and poloidal
   fields in stably stratified stars.

   Of course, the specific entropy $s$ and the proton fraction $Y$ are not perfectly
   conserved quantities within each fluid element, but can be changed by dissipative
   processes, discussed in \S~3: In the case of $s$, through heat diffusion \citep{Parker74}, in the
   case of $Y$, by (direct or inverse) beta decays or by ambipolar
   diffusion (motion of charged relative to neutral particles;
   \citealt{Pethick92,GR92,TD96,Hoyos08}).
   Thus, the condition of stable stratification, and with it the
   hypothetically associated stable magnetic equilibrium
   configuration, although excellent approximations on short
   (Alfv\'en-like) time scales, are eroded on the time scales
   of the dissipative processes mentioned above, leading to the decay
   of these structures \citep{GR92}, and perhaps to a sudden loss of
   stability \citep{BS04,BS06,BN06}. In main-sequence
   stars, white dwarfs, and weakly magnetized neutron stars, these appear
   to be too long to act on the stellar life time, but in strongly
   magnetized neutron stars their time scales become shorter, so
   they might plausibly drive magnetic field decay, leading to
   internal heating and to the magnetar phenomenon \citep{TD96,Arras04}.

   In principle, the Hall drift might also play a role in the
   field evolution in neutron stars
   \citep{Jones88,Urpin,GR92,R05,R07a,Pons}. However, its time
   scale in neutron star cores tends to be somewhat longer than
   those of the other processes considered here \citep{GR92}.
   Moreover, in an axially
   symmetric, equilibrium magnetic field configuration, the effect
   of the Hall drift is exactly cancelled by bulk fluid motions
   \citep{RT08}, so we do not take it into account. For simplicity,
   we also refrain from discussing the role of
   the solid crust of the neutron star, as well as the effects of
   superconductivity and superfluidity, which alter the
   magnetic stresses \citep{Easson77,Akgun08} as well as the
   dissipative processes. We also
   ignore the process of initial set-up of the magnetic
   equilibrium, which might be a highly dynamical process
   involving differential rotation and possibly a dynamo
   \citep{TD93,Spruit02}, but
   concentrate on the properties imposed by the equilibrium and
   stability conditions and on the long-term evolution of the
   field.

   A concise summary of our conclusions is given in \S~4.
   Parts of this discussion have already been given
   elsewhere \citep{R07b,R08}.


\section{Magnetic equilibria and stable stratification}
\label{sec:equilibria}

\subsection{Force balance}\label{sec:force}
   In a conducting, fluid star, a general MHD equilibrium is set
   by the condition that the net force on the fluid vanishes
   everywhere, i.~e.
   \begin{equation}\label{equilibrium}
       {\bf f}_B+{\bf f}_F=0,
   \end{equation}
   where
   \begin{equation}\label{Lorentz}
       {\bf f}_B\equiv{1\over c}{\bf j}\times{\bf B}
   \end{equation}
   is the magnetic (Lorentz) force per unit volume, written in terms
   of the magnetic field, ${\bf B}$, and its associated current density,
   ${\bf j}=(c/4\pi)\nabla\times{\bf B}$, and
   \begin{equation}\label{fluid}
       {\bf f}_F\equiv-\nabla P-\rho\nabla\psi
   \end{equation}
   is the fluid force, which depends on its pressure, $P$, density
   $\rho$, and gravitational potential, $\psi$.

   In all the stars of interest, the fluid is non-barotropic, i.~e.
   the density is not a function of pressure only, but depends on
   an additional, non-trivial variable $X$, which is conserved on dynamical
   (sound or Alfv\'en wave crossing) time
   scales: specific entropy ($X=s$) in the case of upper main sequence stars and
   white dwarfs, and the fraction of protons ($X=Y$) or other minor constituent
   particles required by beta equilibrium
   in the case of neutron stars \citep{Pethick92,RG92}.

   As shown in equation (\ref{beta}), the fluid pressure is much higher than
   the magnetic pressure, so we take the point of view that the magnetic forces create
   only a slight perturbation to the background hydrostatic equilibrium state
   (${\bf f}_F=0$) the star would have in their absence. In addition, we invoke
   the standard
   Cowling approximation of neglecting perturbations to the gravitational
   potential, also used in the simulations of Braithwaite and
   collaborators.

   We do not assume that the unperturbed
   star is spherically symmetric, so our arguments can also be
   applied to stars that are uniformly rotating, in which case
   $\psi$ has to be interpreted as the effective potential, also
   including centrifugal effects. However, we ignore the effects
   of meridional circulation. The time
   scale for this process, due to the interaction of stellar rotation and
   internal heat flow, is
   $t_\mathrm{circ}\sim(\Omega_K/\Omega)^2t_\mathrm{KH}$, where
   $\Omega$ is the stellar rotation rate, $\Omega_K$ is its maximum
   (Keplerian or ``break-up'') value, and $t_\mathrm{KH}$ is the
   (Kelvin-Helmholtz) time scale required to radiate away the
   thermal energy content of the star. For main sequence stars,
   $t_\mathrm{KH}$ is substantially shorter than their
   main-sequence life time, so meridional circulation can modify
   the magnetic field structure of sufficiently fast rotators,
   $P=2\pi/\Omega<5\mathrm{d}$ \citep{Moss84,Moss90,MestelBook}, to which our
   analysis will therefore not apply. For white
   dwarfs, $t_\mathrm{KH}$ is their cooling time, i.~e.
   essentially their age, so meridional circulation is unimportant
   unless they rotate near break-up \citep{Tassoul83}. In
   none of these cases, the magnetic field is expected to be
   strong enough to have a substantial influence on the pattern or
   time scale of meridional circulation. For neutron stars,
   since the main
   source of stratification is not entropy but chemical
   composition \citep{RG92}, meridional circulation will not occur at all.

   Thus, we write the fluid force in terms of the Eulerian
   perturbations (changes at fixed spatial positions) of density
   and pressure, $\delta\rho$ and $\delta P$, respectively, as
   \begin{equation}\label{fluid_pert}
       {\bf f}_F=-\nabla\delta P-\delta\rho\nabla\psi.
   \end{equation}
   The perturbations can be viewed as being produced by a displacement
   field $\xiv$, which allows us to introduce Lagrangian
   perturbations (comparing the variables in the same fluid
   element as it gets displaced) $\Delta\rho$, $\Delta P$, formally
   related to the Eulerian perturbations by
   \begin{equation}\label{Lagrangian}
       \Delta\equiv\delta+\xiv\cdot\nabla,
   \end{equation}
   where the gradient operator acts on the corresponding unperturbed ``background''
   quantity, $\rho_b$ or $P_b$.

   If the displacement is fast enough, it can be taken to
   conserve specific entropy and chemical composition, so the Lagrangian perturbations
   are related by the adiabatic index
   \begin{equation}\label{adiabatic}
       \gamma_p\equiv{\Delta\ln P\over\Delta\ln\rho}
       =\left(\partial\ln P\over\partial\ln\rho\right)_X,
   \end{equation}
   which is generally different from the analogous quantity characterizing the hydrostatic
   equilibrium profile of the star,
   \begin{equation}\label{background}
       \gamma_b\equiv{d\ln P_b\over d\ln\rho_b}
       =\left(\partial\ln P\over\partial\ln\rho\right)_X
       +\left(\partial\ln P\over\partial\ln X\right)_\rho{d\ln X\over d\ln\rho_b}.
   \end{equation}

   Then, also using the condition of mass conservation,
   $\Delta\rho=-\rho_b\nabla\cdot\xiv$,
   equation~(\ref{fluid_pert}) can be manipulated into the form
   \begin{equation}\label{buoyancy}
       {\bf f}_F=-\rho_b\nabla\left(\delta P\over\rho_b\right)
       +\left({\gamma_p\over\gamma_b}-1\right)\Delta\rho\nabla\psi,
   \end{equation}
   where the first term would be present as well in a barotropic,
   homogeneous fluid, whereas the second accounts for buoyancy
   effects. The latter is stabilizing if $\gamma_p>\gamma_b$ and
   destabilizing in the opposite case.

   In upper main sequence stars, the fluid is a classical,
   monatomic ideal gas with $\gamma_p=5/3$, with their
   radiative envelopes well described by $\gamma_b\approx 4/3$
   \citep{MacGregor}, so $\gamma_p/\gamma_b-1\approx
   1/4$. In white dwarfs, the electrons are highly degenerate ($kT\ll E_{Fe}$,
   where $k$ is Boltzmann's constant, $T=T_7\times 10^7\mathrm{K}$ is the interior temperature, and $E_{Fe}$
   is the electron Fermi energy, not including the relativistic rest-mass term, $mc^2$) and
   dominate the pressure, but the entropy is contained in the
   ions, so $\gamma_p/\gamma_b-1\sim kT/ZE_F\sim T_7/500$. Finally,
   in the case of neutron stars, entropy becomes negligible
   a few seconds after their birth, but they remain stably stratified due
   to the density-dependent proton fraction,
   $\gamma_p/\gamma_b-1\sim Y\sim\mathrm{few}~\%$
   \citep{RG92,Lai94,R01a}.

   Eq.~(\ref{buoyancy}) shows that, in a stably stratified fluid (with $\gamma_p>\gamma_b$), the
   fluid force has two parts that are determined by two independent, scalar functions,
   e.~g. $\delta P$ and $\Delta\rho$, which give the fluid a
   greater freedom to balance the magnetic force than it would
   have in the barotropic case ($\gamma_p=\gamma_b$).
   It should be noted that, if the buoyancy term in
   eq.~(\ref{buoyancy}) is crucial to
   balance a particular field configuration of characteristic
   length scale comparable to the stellar radius,
   the characteristic field strength is constrained by
   \begin{eqnarray}\label{maxfield}
   |\mathbf{f}_B|&\sim&{B^2\over 8\pi
   R}\sim\left({\gamma_p\over\gamma_b}-1\right)|\Delta\rho\nabla\psi|\nonumber \\
   &\lsim&\left({\gamma_p\over\gamma_b}-1\right)\rho|\nabla\psi|
   \approx\left({\gamma_p\over\gamma_b}-1\right)|\nabla P|.
   \end{eqnarray}
   Thus, its maximum value is not set by the condition $\beta\gsim 1$
   (with $\beta$ defined in eq. \ref{beta}),
   but rather by the more restrictive
   \begin{equation}\label{limit}
   \beta(\gamma_p/\gamma_b-1)\gsim 1,
   \end{equation}
   yielding maximum allowed field strengths of $\sim 10^8~\mathrm{G}$ for Ap/Bp stars,
   $\sim 10^{11}T_7^{1/2}~\mathrm{G}$ for white dwarfs, and a few times $10^{17}~\mathrm{G}$
   for neutron stars, all of which still substantially exceed the observationally inferred
   fields in these stars.


\subsection{Hierarchy of equilibria and variational principles}
\label{sec:hierarchy}

Given this physical background, we now explore the more
mathematical issue of how magnetic equilibria obtained under
progressively more stringent (and, for our purposes, more
realistic) constraints can be represented by constrained
stationary points of the magnetic energy. We note that, contrary
to the previous section, where ``perturbations'' were deviations
from the non-magnetic, hydrostatic equilibrium caused by the
magnetic field, here perturbations are taken with respect to
successive, \emph{magnetic} equilibria.

\subsubsection{Field-free:}\label{sec:field-free}
Consider the total magnetic energy within a fixed volume,
$U_B=\int_\mathcal V B^2d\mathcal V/(8\pi).$ The only way to
obtain $\delta U_B=0$ under a weak, but otherwise \emph{arbitrary}
magnetic field variation is to have $\mathbf B=0$ everywhere. This
is the absolute minimum of the magnetic energy, and it is
eventually obtained in a star placed in vacuum, without external
fields, and in which a sufficiently effective dissipation
mechanism (such as resistive diffusion) is active.

\subsubsection{Current-free:}\label{sec:current-free}
Of course, $\delta\mathbf B$ is not fully
arbitrary, but must be divergenceless, so we now consider the
slightly more restricted case $\delta\mathbf
B=\nabla\times\delta\mathbf A$. Now,
\begin{eqnarray}
\delta U_B&=&{1\over 4\pi}\int_\mathcal V \mathbf
B\cdot\nabla\times\delta\mathbf A~d\mathcal V\nonumber \\
&=&{1\over 4\pi}\left[\int_\mathcal V \nabla\cdot(\delta\mathbf
A\times\mathbf B)~d\mathcal V+\int_\mathcal V \delta\mathbf
A\cdot\nabla\times\mathbf B~d\mathcal V\right].\label{A}
\end{eqnarray}
In the last result, the first integral can be made to vanish by
imposing appropriate boundary conditions on the surface, for
example that the magnetic flux through any surface element is
fixed (${\mathbf{\hat n}}\cdot\delta\mathbf B=0$, where
${\mathbf{\hat n}}$ is the outward unit normal), which implies
that ${\mathbf{\hat n}}\times\delta\mathbf A=0$. The second term
can only be made to vanish for arbitrary $\delta\mathbf A$ if the
electric current density vanishes everywhere in the volume,
$\mathbf j=c\nabla\times\mathbf B/(4\pi)=0.$ In order to have a
finite magnetic field in a volume containing no currents, there
must be source currents in some neighboring volume. Therefore, a
finite stellar magnetic field cannot be current-free everywhere.

\subsubsection{Force-free:}\label{sec:force-free}
In stars, dissipative processes such as the resistive diffusion of
magnetic flux are slow (see \S~\ref{sec:evolution} for details)
and therefore often negligible. In this context, the only possible
perturbations of the magnetic field are those which can be
produced by plasma displacements, $\delta\mathbf
A=\xiv\times\mathbf B$, with $\xiv$ an arbitrary vector field. In
this case, neglecting the divergence term of eq.~(\ref{A}),
$\delta U_B=-\int_\mathcal V~\xiv\cdot({\mathbf j\times\mathbf
B/c})~d\mathcal V$, so the stationarity of the magnetic energy
implies the vanishing of the Lorentz force, ${\mathbf
j\times\mathbf B/c}=0$. This case is relevant in very diffuse
plasmas such as those in stellar magnetospheres, where the
conductivity is high, but gas pressures and densities are low
($\beta\ll 1$), so the dynamics is dominated by the magnetic
field. However, as shown in Appendix~\ref{sec:forcefree}, it is
not possible to have magnetic field structures that are force-free
everywhere in a star, unless it is confined by (unrealistic)
surface forces.

\subsubsection{Force balance:}\label{sec:forcebalance}
As we saw in \S~\ref{sec:force}, the plasma inside stars has
$\beta\gg 1$, so the fluid forces, due to pressure and gravity,
can by no means be neglected. The total energy perturbation (with
respect to some hydrostatic or hydromagnetic equilibrium state)
caused by an arbitrary fluid displacement field $\xiv$ can be
calculated by integrating the work per unit volume $-(1/2)\mathbf
f\cdot\xiv$ done against the forces of Eqs.~(\ref{Lorentz}) and
(\ref{fluid_pert}) in order to build up this displacement field.
In the absence of a magnetic field, the result can be written as
\begin{eqnarray}
\delta U=\int{1\over 2\gamma_pP_b}\left[\delta
P^2+\left({\gamma_p\over\gamma_b}-1\right)(\xiv\cdot\nabla
P_b)^2\right]~d\mathcal V, \label{hydro_energy}
\end{eqnarray}
where we ignored a surface term that vanishes for appropriate
boundary conditions. Clearly, the $\delta P^2$ term is always
positive, while the other term is positive for a stably stratified
fluid ($\gamma_p>\gamma_b$), zero for a barotropic fluid
($\gamma_p=\gamma_b$), and negative for a convectively unstable
fluid ($\gamma_p<\gamma_b$). Clearly, there are no unstable
(negative-energy) perturbations if $\gamma_p\geq\gamma_b$. In a
stable equilibrium, one must have vanishing Eulerian pressure
perturbations, $\delta
P=-\gamma_pP_b\nabla\cdot\xiv-\xiv\cdot\nabla P_b=0$, and, if the
fluid is stably stratified, also no ``vertical'' displacements,
$\xiv\cdot\nabla P_b=0$. All displacements satisfying these
conditions produce neutrally stable perturbations. We expect these
conditions to still be approximately satisfied in equilibria that
involve a weak magnetic field.

When a magnetic field is introduced, the total energy perturbation
becomes more complicated,
\begin{eqnarray}
\delta U=\int{\Huge[}{\delta\mathbf B^2\over 8\pi}-{1\over
2c}\mathbf j\cdot\delta\mathbf B\times\xiv+\gamma_pP(\nabla\cdot\xiv)^2\nonumber\\
+(\xiv\cdot\nabla
P)\nabla\cdot\xiv-(\xiv\cdot\nabla\psi)\nabla\cdot(\rho\xiv){\Huge]}~d\mathcal
V \label{fullenergy}
\end{eqnarray}
\citep{Bernstein58}, and negative-energy perturbations exist for
some field configurations, even if $\gamma_p\geq\gamma_b$
\citep{Tayler,FlowersRuderman}. However, since these instabilities
are caused by the magnetic field, we expect the quantities
$|\delta P|$ and (if the fluid is stably stratified)
$|\xiv\cdot\nabla P|$ to be small enough to keep the fluid force
perturbation $\delta\mathbf f_F=-\nabla\delta
P-\delta\rho\nabla\psi$ not much larger than the magnetic force
perturbation $\delta\mathbf f_B=(\delta\mathbf j\times\mathbf
B+\mathbf j\times\delta\mathbf B)/c$. This requires
$|\nabla\cdot(\rho\xiv)|\lsim\rho|\xiv|/(\beta\ell)$ for
barotropic and stably stratified fluids, and
$|\xi_r|\lsim|\xiv|/\beta$ only for the latter. Since the
non-magnetic terms in Eq.~(\ref{fullenergy}) are quadratic in
these quantities, they will be smaller than the magnetic terms,
therefore it seems reasonable to consider only magnetic energy
perturbations, subject to the conditions $\nabla\cdot(\rho\xiv)=0$
for barotropic and stably stratified fluids, and additionally
$\xiv\cdot\nabla\psi=0$ for the latter. Note that this bypasses
some instabilities that originate in particular regions where the
fluid forces are weak, such as near the center of the star
\citep{Tayler}.

\paragraph{Barotropic fluid:} Thus, for a \emph{barotropic} fluid with high $\beta$,
we impose the constraint $\delta
P/\gamma_p=\delta\rho=-\nabla\cdot(\rho\xiv)=0$, therefore
$\xiv=(1/\rho)\nabla\times\mathbf a$, where $\mathbf a$ is an
arbitrary vector field. The magnetic energy perturbation becomes
\begin{equation}
\delta U_B=\int_\mathcal V~\nabla\cdot\left[(\mathbf
j\times\mathbf B)\times\mathbf a\over\rho c\right]-\int_\mathcal
V~\mathbf a\cdot\nabla\times\left(\mathbf j\times\mathbf
B\over\rho c\right).
\end{equation}
Consistent with the condition that $\delta\rho=0$ everywhere, we
require ${\mathbf{\hat n}}\cdot\xiv=0$ on the surface of the star,
implying ${\mathbf{\hat n}}\times\mathbf a=0$, which makes the
first integral vanish. The vanishing of the second for arbitrary
$\mathbf a$ requires
\begin{equation}\label{curlfree}
\nabla\times\left(\mathbf j\times\mathbf B\over\rho c\right)=0,
\end{equation}
i.~e. that the Lorentz force per unit mass be a gradient
(consistent with it having to be balanced by the first term in
eq.~[\ref{buoyancy}]). This is the case considered in most
explicit descriptions of neutron star magnetic fields so far
\citep{Tomimura05,Yoshida06,Haskell08,Akgun08,Kiuchi08}.

\paragraph{Stably stratified fluid:}\label{sec:stratified}
Finally, in the strongly non-barotropic, \emph{stably stratified}
case (the most realistic, according to our discussion in
\S~\ref{sec:force}), vertical motions are strongly suppressed, so
$\xiv$ also has to be tangent to gravitational equipotential
surfaces, which is equivalent to requiring that $\mathbf
a=f\nabla\psi$, where $f$ is an arbitrary scalar function. Now
demanding that $\delta U_B=0$ for arbitrary $f$, we obtain
\begin{equation}
\nabla\psi\cdot\nabla\times\left(\mathbf j\times\mathbf B\over\rho
c\right)=0,
\end{equation}
a weaker condition than eq.~(\ref{curlfree}), that is satisfied if
\begin{equation}
{\mathbf j\times\mathbf B\over\rho c}=\nabla\mu+\nu\nabla\psi,
\end{equation}
where $\mu$ and $\nu$ are arbitrary scalar functions, consistent
with eq.~(\ref{buoyancy}).

It is clear that both the limit of a very diffuse plasma
($\beta\ll 1$) applicable to \S~\ref{sec:force-free} and the very
dense plasma ($\beta\gg 1$) of the present section are
idealizations. A more rigorous description would minimize the
total energy of the star, including internal and gravitational
energies in addition to the magnetic energy, and not impose the
additional conditions of this section on the displacement field
$\xiv$. Moreover, the Tayler instability (\citealt{Tayler}; see
also \S~\ref{sec:axisymmetry} of the present paper), although
active for arbitrarily high values of $\beta$, requires to account
explicitly for these other contributions to the energy, and
consistently to relax the constraints on $\xiv$. For our purposes,
the idealized description given here appears to be sufficient.

\subsubsection{A note on helicity conservation:}
A variation of the magnetic helicity, $\mathcal
H\equiv\int_\mathcal V \mathbf A\cdot\mathbf B~d\mathcal V$, can
be written, aside from a surface term \citep{Spruit08}, as
$\delta\mathcal H=2\int_\mathcal V \mathbf B\cdot\delta\mathbf
A~d\mathcal V$, so (for appropriate boundary conditions) helicity
is automatically conserved if $\delta\mathbf A=\xiv\times\mathbf
B$. If we initially allow for an arbitrary $\delta\mathbf A$, but
then search for a stationary point of $U_B$ at fixed $\mathcal H$
(as done by \citealt{Woltjer} and more recently by
\citealt{Broderick08}), we obtain $\mathbf j=\alpha\mathbf B$,
where $\alpha$ is a constant Lagrange multiplier. This condition
is less restrictive than those obtained in \S~\ref{sec:field-free}
and \S~\ref{sec:current-free}, but more restrictive than those of
\S~\ref{sec:force-free} and \S~\ref{sec:forcebalance}. In
particular, the force-free condition of \S~\ref{sec:force-free}
allows for $\mathbf j=\alpha(\mathbf r)\mathbf B$ with $\mathbf
B\cdot\nabla\alpha=0$, so $\alpha$ is constant on a given field
line, but possibly different on different field lines. Thus, the
condition of helicity conservation is most relevant in cases where
$\delta\mathbf A=\xiv\times\mathbf B$ is not exactly satisfied,
i.~e. resistive dissipation allows some motion of magnetic flux
lines with respect to the fluid. Strictly speaking, such motion
does not conserve either energy or helicity. However, helicity is
more strongly dominated by large spatial scales than the magnetic
energy, so small-scale resistive dissipation may conserve the
former to a better approximation than the latter \citep{Field86}.


\subsection{Toy model: a thin flux tube} \label{sec:fluxtube}

As a basis for later, educated guesses about the stability and
evolution of MHD equilibria in stars, we examine the stability of
a thin, azimuthal torus of cross section $A$ lying in the
equatorial plane of the star, at a distance $r\ (\gg\sqrt{A})$
from the center, and containing a weak, roughly uniform azimuthal
magnetic field $B\ (\ll\sqrt{8\pi P_b})$. For a general discussion
of the properties of thin magnetic flux tubes, see
\citet{Parker79} and references therein.

In order to be in equilibrium, the forces across the flux tube's
cross section must balance, which requires the fluid pressure
inside to be lower by $\delta P=-B^2/(8\pi)$. This is achieved on
the very short Alfv\'en crossing time $\sim A^{1/2}/v_A$, where
$v_A=B/\sqrt{4\pi\rho_b}$ is the Alfv\'en speed inside the flux
tube. On a longer time $\sim r/v_A$, the net forces on each
section of the flux tube must also come into balance. Its tension,
$T=AB^2/(4\pi)$ \citep{Parker79}, causes a radial force per unit
length $f_T=-AB^2/(4\pi r)$ that tends to contract the flux tube.
On the other hand, if the entropy and composition of the matter
inside and outside the flux tube are the same, the mass density
inside will be lower than outside, $\delta\rho/\rho_b=\delta
P/(\gamma_pP_b)<0$, causing a radially outward buoyancy force per
unit length, $f_g=-\delta\rho g$, where $g=|\nabla\psi|$ is the
gravitational acceleration.

We take the point of view that the flux tube is initially placed
at a radius $r$ where the matter outside has the same composition
and entropy as inside, and then allowed to displace to $r+\xi_r$,
enforcing $\delta P=-B^2/(8\pi)$ at each point, while the net
force $f_{net}=f_T+f_g$ controls the radial motion. Using the
notation of \S~\ref{sec:equilibria}, we note that
\begin{equation}\label{drop}
\delta\rho={\rho_b\over \gamma_p P_b}\left[\delta
P+\left({\gamma_p\over\gamma_b}-1\right)\rho_b g\xi_r\right],
\end{equation}
so the net force per unit length can be written as
\begin{equation}\label{netforce}
f_{net}={A\over\gamma_pH}\left[\left(1-{2\gamma_pH\over r}\right)
{B^2\over8\pi}-\left({\gamma_p\over\gamma_b}-1\right)\rho_b
g\xi_r\right],
\end{equation}
where the pressure scale height $H\equiv P_b/|\nabla
P_b|=P_b/(\rho_b g)$. The term proportional to $\xi_r$ accounts
for the stratification of the fluid, and is manifestly stabilizing
(force opposing displacement) if $\gamma_p>\gamma_b$ and
destabilizing (force reinforcing displacement) in the opposite
case, while it vanishes for a barotropic fluid, for which
$\gamma_p=\gamma_b$. On the other hand, the term proportional to
$B^2$ is the force on the undisplaced flux tube, which will cause
an inward displacement if $r<2\gamma_pH$ and an outward
displacement in the opposite case, whereas $r=2\gamma_pH$
corresponds to an unstable equilibrium point.

In a stably stratified fluid ($\gamma_p>\gamma_b$), an equilibrium
will be reached at the displacement
\begin{equation}\label{xi_eq}
\xi_r={B^2\over 8\pi\rho_b
g}{1-2\gamma_pH/r\over\gamma_p/\gamma_b-1}.
\end{equation}
Clearly, this equilibrium is stable with respect to radial,
azimuthally symmetric displacements. However, it is intuitive that
the flux tube could contract towards the axis by moving away from
the equatorial plane, roughly on a sphere of radius $r$. This
motion would be driven by the tension, without being opposed by
the buoyancy force. It could only be prevented by an additional,
poloidal magnetic field, which can either be enclosed by the
toroidal flux tube under consideration or be present in the form
of a twist of the magnetic field in the tube.

In all other cases ($\gamma_p\leq\gamma_b$), including that of a
barotropic fluid ($\gamma_p=\gamma_b$), there will be no
equilibrium except at $r=2\gamma_pH$, and the flux tube will
either expand (if $r>2\gamma_pH$) or contract (if $r<2\gamma_pH$)
indefinitely, at a speed determined by the fluid drag force
\citep{Parker74}.

This simple example suggests that, in the general case, the
stratification of the fluid is likely to play an important role in
determining the structure of magnetic equilibria, in the sense
that there should be a much wider variety of possible equilibria
in a stably stratified fluid than in a barotropic one.


\subsection{Axially symmetric equilibria} \label{sec:axisymmetry}

   The stable equilibria found by Braithwaite and collaborators \citep{BS04,BS06,BN06}
   can be described ideally as axially symmetric
   (but see \citealt{Braithwaite08} for highly asymmetric equilibria),
   involving two distinct regions: a thick
   torus fully contained in the star and containing a twisted
   toroidal-poloidal field combination, and the rest of space, containing a purely
   poloidal field that goes through the hole in the torus, and
   closing outside the star, in this way giving the external field
   an essentially dipolar appearance. It had long been speculated
   that such stable configurations might exist, but this has never
   been
   confirmed analytically (see \citealt{BN06} for a
   discussion of earlier work).

   For this reason, here we assume axial symmetry, allowing for
   both poloidal and toroidal field components. In this case, all
   the fluid variables
   depend only on two of the
   three cylindrical coordinates, $\varpi$ and $z$. The most general,
   axially symmetric magnetic field can be
decomposed into a toroidal component
\begin{equation}\label{toroidal}
{\bf B}_T={\cal B}(\varpi,z)\nabla\phi
\end{equation}
and a poloidal component
\begin{equation}\label{poloidal}
{\bf B}_P=\nabla{\cal A}(\varpi,z)\times\nabla\phi
\end{equation}
\citep{Chandrasekhar56}. This decomposition makes it explicit that
the field depends only on two scalar functions, ${\cal B}$ and
${\cal A}$, and explicitly satisfies the condition of zero
divergence independently for both components. As in \citet{R07a},
we choose to write it in terms of the gradient of the azimuthal
angle, $\nabla\phi=\hat\phi/\varpi$, instead of the unit vector,
$\hat\phi$, in order to make easy use of the identity
$\nabla\times\nabla\phi=0$. For reference, we also write the
toroidal and poloidal components of the current,
\begin{eqnarray}
\label{torcurrent}{\bf j}_T=&{\displaystyle{c\over
4\pi}}\nabla\times{\bf B}_P
=&-{\displaystyle{c\over 4\pi}}\Delta^*{\cal A}~\nabla\phi,\\
\label{polcurrent}{\bf j}_P=&{\displaystyle{c\over
4\pi}}\nabla\times{\bf B}_T=&{\displaystyle{c\over
4\pi}}\nabla{\cal B}\times\nabla\phi,
\end{eqnarray}
where $\Delta^*\equiv \varpi^2\nabla\cdot(\varpi^{-2}\nabla)$ is
sometimes known as the ``Grad-Shafranov operator'', although it
appears to have been first introduced by \citet{Lust54}.
Eqs.~(\ref{toroidal}) through (\ref{polcurrent}) show that the
magnetic field lines lie on the surfaces ${\cal
A}=\mathrm{constant}$, while the current lines lie on surfaces
${\cal B}=\mathrm{constant}$. If both $\mathcal{A}$ and
$\mathcal{B}$ are taken to be zero on the symmetry axis, then
$2\pi{\cal A}$ is the poloidal flux enclosed by a given surface
${\cal A}=\mathrm{constant}$, whereas $c{\cal B}/2$ is the total
current enclosed by the corresponding surface (see also
\citealt{Kulsrud}, \S~4.9).

   In axial symmetry, the gradients in eqs.~(\ref{fluid}) and (\ref{fluid_pert}) do not
   have an azimuthal component, and therefore eq.~(\ref{equilibrium})
   requires ${\bf f}_B\cdot\hat\phi=0$, or equivalently
   \begin{equation}\label{noforce}
   {\bf j}_P\times{\bf B}_P=0,
   \end{equation}
   i.~e. ${\bf j}_P$ and ${\bf B}_P$ must be parallel to each other
   everywhere. (Note that ${\bf j}_T$ and ${\bf B}_T$ are
   \emph{always} parallel.) In terms of the scalar functions defined above,
   \begin{equation}\label{scalarnoforce}
   \nabla{\cal A}\times\nabla{\cal B}=0,
   \end{equation}
   i.~e. the surfaces ${\cal A}=\mathrm{constant}$ and ${\cal
   B}=\mathrm{constant}$ coincide, making it possible to write one
   of these functions in terms of the other, e.~g. ${\cal
   B}={\cal B}({\cal A})$ \citep{Chandrasekhar56}. When this condition is satisfied,
   \begin{equation}\label{Lorentzpoloidal}
   {\bf f}_B=-{1\over 4\pi \varpi^2}\left(\Delta^*{\cal A}+{\cal B}{d{\cal
   B}\over d{\cal A}}\right)\nabla{\cal A},
   \end{equation}
   with only two vector components (in the $\varpi-z$ plane). Thus, for any choice of
   the functions ${\cal A}(\varpi,z)$ and ${\cal B}({\cal A})$ whose magnitude is small
   enough to satisfy eq.~(\ref{limit}), it should be possible to find independent scalar
   functions $\delta P$ and $\Delta\rho$ in eq.~(\ref{fluid_pert}) that
   yield an equilibrium state. Thus, as already realized by \citet{Mestel56},
   \emph{any (weak) axially symmetric field satisfying
   eq.~(\ref{noforce}) corresponds to a magnetostatic equilibrium in a stably
   stratified fluid.}

   The possible equilibria are much more restricted in the barotropic case, in which
   the stabilizing $\Delta\rho$ term in eq.~(\ref{fluid_pert}) vanishes and
   the fluid force depends on a single scalar function $h\equiv\delta
   P/\rho_b$. Using this, together with
   eq.~(\ref{Lorentzpoloidal}), in the force-balance equation
   (\ref{equilibrium}), one finds that $\nabla h$ is parallel to
   $\nabla{\cal A}$, so $h=h({\cal A})$, and eventually one
   obtains the popular Grad-Shafranov equation (e.~g. \citealt{Kulsrud}, \S~4.9),
   \begin{equation}\label{GS}
   \Delta^*{\cal A}+{\cal B}{d{\cal B}\over d{\cal A}}=-4\pi
   \varpi^2\rho_b{dh\over d{\cal A}},
   \end{equation}
   which is often assumed to characterize stellar magnetic fields
   \citep{Tomimura05,Yoshida06,Haskell08,Akgun08,Kiuchi08}. We
   emphasize that, in all the stars of interest here, the fluid is
   \emph{not} barotropic, but stably stratified, with stabilizing
   buoyancy forces much stronger than the
   Lorentz forces, so the magnetic equilibria are \emph{not}
   required to satisfy eq.~(\ref{GS}), but only the condition
   contained in eqs.~(\ref{noforce}) and (\ref{scalarnoforce}).

   Of course, the existence of an equilibrium does not guarantee its
   stability, which is clearly illustrated by the two simplest
   cases of purely toroidal and purely poloidal fields, for which
   there are equilibria, which however are always unstable to
   non-axisymmetric perturbations. For a
   purely toroidal field, flux rings can shift with respect to
   each other on spherical surfaces, in this way reducing the
   total energy of the configuration
   \citep{Tayler}. For a purely poloidal field, one can imagine
   cutting the star along a plane parallel to the symmetry axis
   and rotating one half with respect to the other, eliminating
   the dipole moment and reducing the energy of the external field,
   without changing the internal one \citep{FlowersRuderman}.

   In the long-lived configurations found numerically  by
   Braithwaite and collaborators,
   it is clear that the toroidal and poloidal field components
   might stabilize each other against both kinds of
   instabilities mentioned in the previous paragraph. We can add
   here, based on the previous discussion, that the toroid of
   twisted field lines can be seen as a collection of nested,
   toroidal surfaces on which lie both the magnetic field lines and the
   current density lines (although their winding angles are
   generally different). As a consequence, the configuration has no
   toroidal Lorentz force component, although it generally does have
   poloidal components that are balanced by a pressure gradient and
   gravity.

   We stress that Braithwaite's and his colleagues' simulations
   considered a single-fluid, \emph{stably stratified} star.
   We can view
   the toroid, at least qualitatively, as a thick version of the
   thin flux tube discussed in \S~\ref{sec:fluxtube}. It is impeded
   from contracting onto the axis by the presence of the poloidal
   flux going through it, as well as by the material between
   the torus and the axis. In the stably
   stratified star, the matter inside the toroid may have a
   slightly different entropy or composition than outside,
   cancelling its tendency to radial expansion due to buoyancy.
   However, if the star were not stably stratified (or this
   stratification could be overcome somehow; see \S~\ref{sec:evolution}), then a
   toroid sufficiently close to the surface would tend to rise and
   eventually move out of the star. If instead it were deep inside
   the star, it would naturally tend to contract due to tension,
   but be impeded from closing on its center by the poloidal flux.
   However, in this case, a small displacement of the whole
   configuration along the axis would cause a net buoyancy force
   that would tend to move it out of the star, along the axis.
   Although it is by no means clear whether this effect leads
   to an instability or instead is quenched by other effects, such
   as the progressive thickening of the toroidal ring or the
   material trapped by the poloidal field, we conjecture that
   stable equilibria occur only in stably
   stratified stars, and not in barotropic ones. If this
   conjecture were correct, it would make the usual search for
   barotropic (Grad-Shafranov) equilibria in fluid stars astrophysically
   meaningless\footnote{They
   might, however, play a role as stable ``Hall equilibria'' in solid neutron
   star crusts \citep{LR09}.}.


\section{Long-term field evolution}
\label{sec:evolution}

From the previous discussion, it becomes natural to suggest that,
as an alternative to the generally slow Ohmic diffusion
\citep{Cowling45,Baym69}, the decay of the magnetic energy may be
promoted by processes that progressively erode the stable
stratification of the stellar matter.

For example, in an entropy-stratified, plane-parallel atmosphere,
a horizontal flux tube with a purely longitudinal magnetic field
$B$ can reach a mechanical equilibrium in which the interior
entropy differs from the external one by $\delta s<0$, so as to
compensate for the fluid pressure difference induced by the
magnetic field, $\delta P=-B^2/(8\pi)<0$, yielding the same mass
density inside as outside, $\delta\rho=(1/c_s^2)[\delta
P-(\partial P/\partial s)_\rho\delta s]=0,$ and thus zero net
buoyancy. However, in this state, the temperature inside the flux
tube is also lower than outside, so heat will stream inwards,
reducing $|\delta s|$ and making the flux tube rise
\citep{Parker74,MacGregor}. This is the main alternative to
resistive diffusion in the case of entropy-stratified stars, i.~e.
upper main sequence stars and white dwarfs.

Similarly, magnetic equilibria in a neutron star rely on a
perturbation of the proton fraction, $\delta Y$, which can be
reduced by two processes \citep{GR92}: \begin{itemize} \item[1)]
direct and inverse beta decays, converting neutrons into charged
particles (protons and electrons) and vice-versa, and \item[2)]
ambipolar diffusion, i.~e. diffusion of charged particles, pushed
by Lorentz forces, with respect to neutral ones.\end{itemize}

In each case, if the magnetic structure was held in place, the
imbalance ($\delta s$ or $\delta Y$) would decay to zero on some
characteristic diffusive or decay timescale $t_c$. However, this
decay corresponds to only a small fraction $\sim B^2/(8\pi
P)=\beta^{-1}$ of the absolute value of the relevant variable ($s$
or $Y$), and therefore is compensated by a similarly small spatial
displacement in the magnetic structure, $\xi_r\sim R/\beta$. A
substantial change in the magnetic structure occurs only on the
much longer time scale $t_B\sim\beta t_c$.

Since these processes and the corresponding $t_c$ differ
substantially from one type of star to another, we now discuss
each type separately.

\subsection{Upper main sequence stars}\label{sec:MS}

The Ohmic dissipation time for a magnetic field in a
non-degenerate star is
\begin{equation}
t_\Omega\sim {\ell^2T^{3/2}\over K}\sim3\times
10^{11}~\mathrm{yr}~\left(\ell\over R_{\sun}\right)^2\left(T\over
10^6~\mathrm{K}\right)^{3/2},
\end{equation}
with Spitzer magnetic diffusivity $\eta=K/T^{3/2}$ and $K\sim
10^{12}~\mathrm{cm^2~s^{-1}~K^{3/2}}$. In order to obtain decay
over the main-sequence lifetime of an A star, $\sim
10^9~\mathrm{yr}$, the characteristic length scale of the magnetic
field configuration would have to be $\ell\sim 0.1~R_{\sun}$,
somewhat smaller than found by \citet{BS04}.

According to the discussion above, the time scale for decay of the
field due to destabilization by heat exchange is $t_B\sim\beta
t_c$, where, in this case, $t_c$ is the heat diffusion time into a
magnetic structure of characteristic scale $\ell$, related to the
Kelvin-Helmholtz time, $t_{KH}$, of the star (of radius $R$) by
$t_c\sim(\ell/R)^2 t_{KH}$; therefore
\begin{eqnarray}
t_B&\sim&\beta\left(\ell\over R\right)^2{GM^2\over RL}\nonumber \\
&\gsim& 10^{14}~\mathrm{yr}~\left(\ell\over R\right)^2\left(R\over
R_{\sun}\right)^{-1}\left(M\over M_{\sun}\right)^2\left(L\over
L_{\sun}\right)^{-1}.
\end{eqnarray}
For realistic numbers, this time scale is comparable or somewhat
longer than the Ohmic time, thus not likely to be relevant for the
star's magnetic evolution.

\subsection{White dwarfs}\label{sec:WD}

In white dwarfs, the same processes are active as in main sequence
envelopes, although modified by the degenerate conditions. The
Ohmic time scale is reduced (factor $\sim 10^{-5}$) by the smaller
length scale, and increased (factor $\sim 10^6$) by the higher
kinetic energy of the electrons (Fermi energy rather than thermal
energy). Thus, again, the resistive decay of a large-scale field
is too slow to play a substantial role in the evolution of these
stars \citep{Wendell}.

Heat diffusion occurs chiefly through transport by the degenerate
electrons, with conductivity
$\kappa={3\pi^3\hbar^3n_ek^2T/(4Ze^4{m_e^*}^2\Lambda)}$, where
$\hbar$ is Planck's constant, $Z$ is the atomic number of the
ions, $e$ is the proton charge, $m_e^*$ is the effective mass of
the electrons (relativistic Fermi energy divided by $c^2$, and
$\Lambda$ is the dimensionless ``Coulomb logarithm'' (see
\citealt{Potekhin99}), which we take $\sim 1$ for the estimates
that follow. Most of the heat is contained in the non-degenerate
ions, whose number density is $n_i=n_e/Z$, so the heat diffusion
time\footnote{For white dwarfs, the diffusion time through scale
$\ell=R$ in the degenerate interior is \emph{not} the
Kelvin-Helmholtz or cooling time, as the bottleneck for the latter
is the conduction through the non-degenerate atmosphere.} through
a scale $\ell$ is
\begin{equation}
t_c\sim {n_ik\ell^2\over\kappa}\sim{4\times 10^7~\mathrm{yr}\over
T_7}\left(\ell\over R\right)^2. \label{WDdiff}
\end{equation}
Imposing the magnetic flux loss time $t_B\sim\beta t_c$ to be
shorter than the cooling time of the star, roughly given by
Mestel's law \citep{Mestel52} as $t_{cool}\sim
10^9~\mathrm{yr}/T_7^{2.5}$, yields the condition
\begin{equation}
{\ell\over R}<{5\over\sqrt{\beta T_7}}. \label{WDcond}
\end{equation}
Since $\beta\geq 10^6$ and the temperature never drops below $\sim
10^5~\mathrm{K}$, only very small-scale magnetic structures, very
different from those found by \citet{BS04}, will be able to decay
by this process.

Unlike the case of neutron stars (\S~\ref{sec:NS}), in known white
dwarfs the thermal energy appears to be always larger than the
magnetic energy, thus the eventual feedback of the magnetic
dissipation on the stellar cooling is negligible.

\subsection{Neutron stars}\label{sec:NS}

Like white dwarfs, neutron stars are passively cooling objects, in
which the progressive decrease of the temperature makes the
reaction rates and transport coefficients (but not the spatial
structure) change with time. In particular, with decreasing
temperature, beta decay rates decrease dramatically, while
collision rates also decrease and thus make particle diffusion
processes proceed more quickly.

In the discussion below, we ignore the possibility of Cooper
pairing of nucleons, which is expected to occur at least in some
parts of the neutron star core and turns neutrons into a
superfluid and protons into a superconductor. This is likely to
have a strong effect on the rates mentioned in the previous
paragraph. However, this effect is difficult to quantify;
therefore we rely on the better-known properties of ``normal''
degenerate matter and leave it to future work to explore the
Cooper-paired analog.

\subsubsection{Direct and inverse beta decays}\label{sec:ENS}

We first consider a hot neutron star (in the neutrino cooling
regime, e.~g. \citealt{Yakovlev}), in which the collision rates
are so high as to effectively bind all particle species together,
but weak interaction processes proceed at non-negligible rates.

For illustration, let us again consider the toy model of a thin,
horizontal magnetic flux tube that is rising due to magnetic
buoyancy through a degenerate gas of neutrons ($n$), protons
($p$), and electrons ($e$). Since the time scale to reach chemical
equilibrium is much longer than any dynamical times, the flux tube
can be considered to be in hydrostatic equilibrium, e.~g. its
internal mass density is equal to that of its surroundings,
$\delta\rho=0$, and its internal fluid pressure is reduced to
compensate for the magnetic pressure, $\delta P=-B^2/(8\pi)$.
These two conditions are only compatible if the fluid inside the
flux tube is not in chemical equilibrium, namely
\begin{equation}
\eta\equiv\mu_n-\mu_p-\mu_e=-\left(\partial\eta\over\partial
P\right)_\rho {B^2\over 8\pi},
\end{equation}
where $\mu_i$ are the chemical potentials of the three particle
species. In order to change $P$ without changing $\rho$, the
composition, here parameterized by the proton or electron
fraction, $Y=n_p/(n_n+n_p)=n_e/(n_n+n_p)$, must change as well.

For a simplified equation of state with non-interacting fermions
(see, e.~g. \citealt{Shapiro}), we show in Appendix
\ref{thermodynamics} that $(\partial P/\partial\eta)_\rho\approx
n_e$, so
\begin{equation}\label{imbalance}
-\eta\approx{B^2\over 8\pi n_e}\sim 2.6~B_{16}^2~\mathrm{keV}\sim
k\times 10^9~B_{16}^2~\mathrm{K},
\end{equation}
where we have assumed small perturbations, $|\eta|\ll\mu_e$, i.~e.
$B\ll(8\pi n_e\mu_e)^{1/2}\sim 4\times 10^{17}~\mathrm{G}$, easily
compatible even with magnetar field strengths. (We took
$n_e\approx 2\times 10^{37}~\mathrm{cm^{-3}}$ for the numerical
estimates.)

In order for the flux tube to move, $|\eta|$ has to decrease by
inverse beta decays, $p+e\to n+\bar\nu_e$, i.~e. by one of the
same processes (direct or modified Urca) that control the cooling
of the star.

In the ``subthermal'' regime \citep{Haensel}, $|\eta|\lsim kT$,
the available phase space for these reactions is determined by the
temperature, and the time scale for the decay of $|\eta|$ is $\sim
10$ times shorter than that for the decrease of $T$ \citep{R95}.
On this time scale, $\eta$ would approach zero if the flux tube
was held at its initial position. What happens is that, as $Y$ is
decreased by the beta decays, the pressure inside the flux tube
increases, the tube expands and rises to find a new hydrostatic
equilibrium in which it continues to be in a slight chemical
imbalance as described by eq.~(\ref{imbalance}). This allows us to
relate the logarithmic changes in the proton fraction inside the
flux tube and the temperature in the star as the latter cools and
the flux tube rises,
\begin{equation}
{d\ln Y\over d\ln T}\sim 10{\eta/Y\over(\partial\eta/\partial
Y)_\rho}\approx 5{B^2\over n\mu_e}.
\end{equation}
A substantial displacement of the flux tube corresponds to this
quantity being $\gsim 1$, i.~e. it requires $B\gsim
(n\mu_e/5)^{1/2}\sim 10^{17}~\mathrm{G}$, stronger than inferred
magnetar fields and dangerously close to violating the linear
limit set above. (In addition, it would require $T\gg
10^{11}~\mathrm{K}$ for consistency with the ``subthermal''
condition.) For weaker fields, the star cools too fast for the
flux tube motion to keep up.

\begin{figure}
\includegraphics[width=13cm]{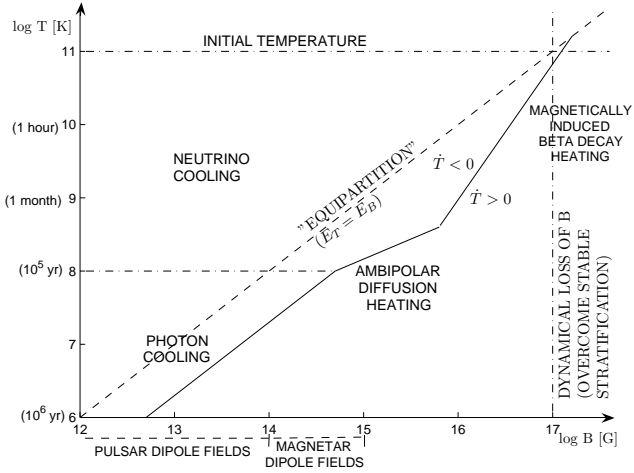} \caption{Magnetic
field -- temperature plane for a non-superfluid neutron star core.
The dot-dashed horizontal lines show the initial temperature (just
after core collapse), and the transition from neutrino-dominated
(modified Urca) to photon-dominated cooling. The dashed diagonal
line corresponds to the equality of magnetic and thermal energy.
Above and to the left of the solid line, the star cools passively,
on the time scales indicated in parenthesis along the vertical
axis, without substantial magnetic field decay, so the evolution
of the star is essentially a downward vertical line. Once the
solid line is reached, magnetic dissipation mechanisms become
important and generate heat that stops the cooling. The subsequent
evolution is expected to be roughly along this line, with
temperature and magnetic field decreasing together, much more
slowly than the passive cooling.}
\end{figure}

On the other hand, in the strongly ``suprathermal'' regime,
$|\eta|\gsim 5 kT$, the induced inverse beta decays leave more
thermal energy in the star than is emitted in the form of
neutrinos, i.~e. in the region in which this chemical imbalance is
present, the Urca processes have a net heating effect \citep{FR05}
and might therefore be able to keep the star warm during a time
long enough for the field to decay \citep{TD96}. This
heating-cooling balance occurs at
\begin{equation}
T_8\sim 2~B_{16}^2.
\end{equation}
On this line (see Fig.~1), the thermal energy in the star,
\begin{equation}
E_T\sim 10^{46}~T_8^2~\mathrm{erg},
\end{equation}
is less than its magnetic energy,
\begin{equation}
E_B\sim 10^{50}~B_{16}^2~\mathrm{erg},
\end{equation}
by a factor
\begin{equation}
E_T/E_B\sim 2\times 10^{-4}~T_8,
\end{equation}
and therefore the cooling process of the star will be delayed by
the inverse of this factor.

\subsubsection{Ambipolar diffusion}\label{sec:LNS}

At somewhat lower temperatures, collision rates are reduced (due
to the increased degeneracy and reduced number of available
quantum states), allowing different particle species to drift with
respect to each other. The Lorentz force only acts directly on the
charged particles (protons, electrons, and perhaps others),
pushing them through the neutrons. The magnetic flux is only
frozen into the charged particle fluid, which moves through the
neutral fluid as fast as the balance of Lorentz force and
collisions allows. If the charged fluid contains only protons and
electrons, it will be barotropic. If additional particle species
are present, it will be stably stratified due to their
density-dependent abundances.

\citet{GR92} decompose the charged particle flux $n_c\vec v_c$ of
ambipolar diffusion into two modes:
\begin{itemize}
\item[1)] Irrotational, with $\nabla\cdot(n_c\vec v_c)\neq 0$ and
$\nabla\times(n_c\vec v_c)=0$, which builds up pressure gradients
in the charged particle fluid, which need to be eliminated by weak
interactions in order for the motion to proceed.

\item[2)] Solenoidal, with $\nabla\cdot(n_c\vec v_c)=0$ and
$\nabla\times(n_c\vec v_c)\neq 0$, corresponding to an
incompressible charged-particle flow, which does not cause
pressure gradients and only needs to overcome the frictional force
due to charged-neutron collisions.
\end{itemize}

The solenoidal mode is analogous to the motion of a barotropic,
incompressible fluid, which should be enough to overcome the
constraints imposed by stable stratification. In a non-superfluid
$npe$ fluid, this mode proceeds on a time scale \citep{GR92}
\begin{equation}
t_\mathrm{ambip}^\mathrm{s}\sim 3\times 10^3{T_8^2\ell_5^2\over
B_{15}^2}~\mathrm{yr},
\end{equation}
causing a magnetic energy dissipation
\begin{equation}
-\dot E_B\sim 0.5\times
10^{44}~{B_{15}^4R_6^3\over\ell_5^2T_8^2}~\mathrm{erg~s^{-1}},
\end{equation}
that can, at sufficiently low temperatures, balance the dominant
cooling luminosity, be it neutrinos (here for the modified Urca
process),
\begin{equation}
L_\nu\sim 3\times 10^{32}T_8^8~\mathrm{erg~s^{-1}},
\end{equation}
or thermal electromagnetic radiation from the stellar surface,
\begin{equation}
L_\gamma\sim 10^{33}~T_8^{2.2}~\mathrm{erg~s^{-1}}.
\end{equation}
The first will happen at $T_8\sim B_{15}^{2/5}$, and the second at
$T_8\sim B_{15}^{0.95}$ (see Fig.~1).

\subsubsection{Neutron star summary}

Strongly magnetized neutron stars appear to be subject to
processes that can erode the stable stratification and therefore
cause an MHD-stable field to decay on time scales shorter than
their observable lifetime. These processes are weak decays, which
are dominant at very high field strengths, and ambipolar
diffusion, at somewhat lower field strengths. In both cases, these
processes become important only once the thermal energy in the
star is substantially less than the magnetic energy, and therefore
the latter acts as a large reservoir that keeps the star hot for
much more than its cooling time in the un-magnetized case (see
also \citealt{TD96,Pons07PRL}). If the field decayed homologously,
the star would evolve following the line of heating-cooling
balance in Fig.~1. In fact, the evolution is likely more complex,
involving loss of stability, followed by abrupt re-arrangements of
the field \citep{BS04}, but these effects should occur roughly on
the heating-cooling balance line.

Of course, neutron stars also have a solid crust, whose elastic
and yielding properties are still highly uncertain. At very high
field strengths, the Lorentz forces will distort the crust, which
might act essentially as a fluid. At lower field strengths, the
crust might act as a valve, controlling the loss of magnetic flux.
The relative importance of the decay mechanisms in the crust (Hall
drift, crust cracking) and core is still unclear, depending on the
uncertain properties of both.


\section{Conclusions}
\label{sec:conclusions}

This paper contains a general discussion of several physical
issues related to the existence of large-scale, coherent magnetic
structures in upper main sequence stars, white dwarfs, and neutron
stars.  The main conclusions are the following:

\begin{itemize}
\item[1)] Magnetic forces in these objects are generally weak
compared to pressure and gravity forces, and their matter is
strongly stratified by entropy or composition gradients. This
means that at least some components of the magnetic forces can
easily be balanced by other forces. Thus, there can be a wide
variety of possible equilibria. These equilibria are not
force-free; in fact, force-free equilibria are not possible in
stars.

\item[2)] If the magnetic structure is axially symmetric, the only
constraint it has to satisfy to be balanced by pressure and
gravity forces is that the azimuthal component of the Lorentz
force must vanish. This means that there must be a set of magnetic
surfaces of toroidal topology containing both the magnetic field
lines and the current flow lines. Since the fluid is not
barotropic, there is no need for the magnetic field to satisfy a
Grad-Shafranov equation.

\item[3)] It is difficult to give general criteria for stability.
However, it is likely that, in a stably stratified star, poloidal
and toroidal field components of similar strength could stabilize
each other. In a barotropic fluid, it is possible that no stable
equilibria exist, as the magnetic field might rise buoyantly and
be lost from the star.

\item[4)] The long-term evolution of the magnetic field is likely
to be governed by dissipative processes that erode the stable
stratification. Heat diffusion in main sequence stars and white
dwarfs appears to be too slow to cause an observable effect over
the life time of these stars. In strongly magnetized neutron
stars, ambipolar diffusion and beta decays might be causing the
magnetic energy release observed in magnetars.
\end{itemize}

\appendix
\section{No force-free fields in stars}\label{sec:forcefree}

Consider the following integral over a volume containing the star
of interest:
\begin{equation}\label{integral}
\int_\mathcal V\mathbf r\cdot(\mathbf j\times\mathbf
B/c)~d\mathcal V=\oint_{\mathcal S(\mathcal V)}r_i T_{ij}
ds_j-\int_\mathcal V T_{ii}d\mathcal V
\end{equation}
(e.~g. \citealt{Kulsrud}, Chapter 4), where the Einstein summation
convention is being used, and the magnetic stress tensor is
\begin{equation}
T_{ij}={B_iB_j\over 4\pi}-{B^2\over 8\pi}\delta_{ij}.
\end{equation}
The last term in eq.~(\ref{integral}) is minus the total magnetic
energy within $\mathcal V$, $U_B=\int_\mathcal V
B^2/(8\pi)~d\mathcal V>0$. The surface integral, taking the
surface to be a sphere of radius $r$, becomes
\begin{equation}
{1\over 8\pi}\oint d\Omega~r^3 B^2\cos(2\beta),
\end{equation}
where $\beta$ is the local angle between $\mathbf B$ and
${\mathbf{\hat r}}$. Outside a star, $B$ falls at least as fast as
$r^{-3}$, so this integral goes to zero as $r\to\infty$. Thus, the
only way to have $\mathbf j\times\mathbf B=0$ everywhere is to
have $U_B=0$, i.~e. $\mathbf B=0$ everywhere (except, perhaps, at
a set of points of measure zero). This means that no magnetic
stars can exist whose field is force-free everywhere. The
``force-free'' configurations of \citet{PerezAzorin06} or
\citet{Broderick08} do not violate this theorem, because they have
current sheets with infinite Lorentz forces on the stellar
surface.

\section{Thermodynamic properties of a degenerate \emph{npe}
fluid}\label{thermodynamics}

Taking the neutrons and protons to be nonrelativistic, the
electrons extremely relativistic, and all highly degenerate, the
total energy density and pressure are
\begin{eqnarray}\label{rhoP}
\rho&=&nmc^2+{3\over 5}(n_n\tilde\mu_n+n_p\tilde\mu_p)+{3\over
4}n_e\mu_e,\nonumber \\
P&=&n^2\left(\partial(\rho/n)\over n\right)_Y={2\over
5}(n_n\tilde\mu_n+n_p\tilde\mu_p)+{1\over 4}n_e\mu_e,
\end{eqnarray}
where $n_n=(1-Y)n$, $n_p=n_e=Yn$, $\mu_e=\hbar
c(3\pi^2n_e)^{1/3}$,
$\tilde\mu_i=\mu_i-mc^2=\hbar^2(3\pi^2n_i)^{2/3}$ for $i=n,p$. The
chemical equilibrium state minimizes the energy per baryon with
respect to $Y$ at fixed $n$, so in equilibrium
$(\partial\rho/\partial Y)_n=0$. Thus, also evaluated at chemical
equilibrium,
\begin{eqnarray}\label{dPdY}
\left(\partial P\over\partial Y\right)_\rho&=&\left(\partial
P\over\partial Y\right)_n=n\left[{2\over 3}(-\mu_n+\mu_p)+{1\over
3}\mu_e\right]\nonumber \\&=&-{n\mu_e\over 3}.
\end{eqnarray}
The chemical imbalance
$\eta=\mu_n-\mu_p-\mu_e=\tilde\mu_n-\tilde\mu_p-\mu_e=-(1/n)(\partial\rho/\partial
Y)_n$ satisfies
\begin{eqnarray}\label{detadY}
\left(\partial\eta\over\partial
Y\right)_\rho&=&\left(\partial\eta\over\partial
Y\right)_n=-{2\over 3}{\tilde\mu_n\over 1-Y}-{2\over
3}{\tilde\mu_p\over Y}-{1\over 3}{\mu_e\over Y}\nonumber \\
&\approx&-{\mu_e\over 3Y}.
\end{eqnarray}
So,
\begin{equation}\label{dPdeta}
\left(\partial P\over\partial\eta\right)_\rho\approx nY=n_e=n_p.
\end{equation}

\begin{acknowledgements}
      The author thanks Stefano Bagnulo, Jon Braithwaite, Peter Goldreich, Swetlana
      Hubrig, Maxim Lyutikov, Friedrich Meyer, and Chris Thompson
      for many stimulating and informative conversations,
      Henk Spruit and an anonymous referee for insightful comments that improved the
      manuscript, and Crist\'obal Petrovich for preparing Fig.~1.
      This work was supported by Proyecto Regular FONDECYT
      1060644 and Proyecto Basal PFB-06/2007.
\end{acknowledgements}

\end{document}